\pdfoutput=1
\documentclass[10pt,letterpaper,twocolumn]{article}
\usepackage{hero-preprint}
\newcommand{\paperauthors}{%
  Zhipeng Yi\textsuperscript{1}, Xinyu Lin\textsuperscript{2},
  Minxuan Shan\textsuperscript{1}, Zipeng Chen\textsuperscript{1},
  Yujiao Zhao\textsuperscript{3}, Chuanyu Xu\textsuperscript{3},
  Wenjie Wang\textsuperscript{1}}
\newcommand{\paperaffiliations}{%
  \textsuperscript{1}University of Science and Technology of China\quad
  \textsuperscript{2}National University of Singapore\quad
  \textsuperscript{3}Alibaba Group}
\newcommand{\paperauthornotes}{}
\newcommand{\papercontact}{%
  \href{mailto:zhipengyi@mail.ustc.edu.cn}{\texttt{zhipengyi@mail.ustc.edu.cn}},
  \href{mailto:xylin1028@gmail.com}{\texttt{xylin1028@gmail.com}},
  \href{mailto:smx574@mail.ustc.edu.cn}{\texttt{smx574@mail.ustc.edu.cn}},
  \href{mailto:zpchen0217@mail.ustc.edu.cn}{\texttt{zpchen0217@mail.ustc.edu.cn}}\\[1pt]
  \href{mailto:joyce.zyj@alibaba-inc.com}{\texttt{joyce.zyj@alibaba-inc.com}},
  \href{mailto:tracy.xcy@taobao.com}{\texttt{tracy.xcy@taobao.com}},
  \href{mailto:wenjiewang96@gmail.com}{\texttt{wenjiewang96@gmail.com}}}
\newcommand{\pdfauthors}{Zhipeng Yi; Xinyu Lin; Minxuan Shan; Zipeng Chen; Yujiao Zhao; Chuanyu Xu; Wenjie Wang}
\newcommand{\institutionlogos}{%
  \raisebox{-0.5\height}{\includegraphics[height=7mm]{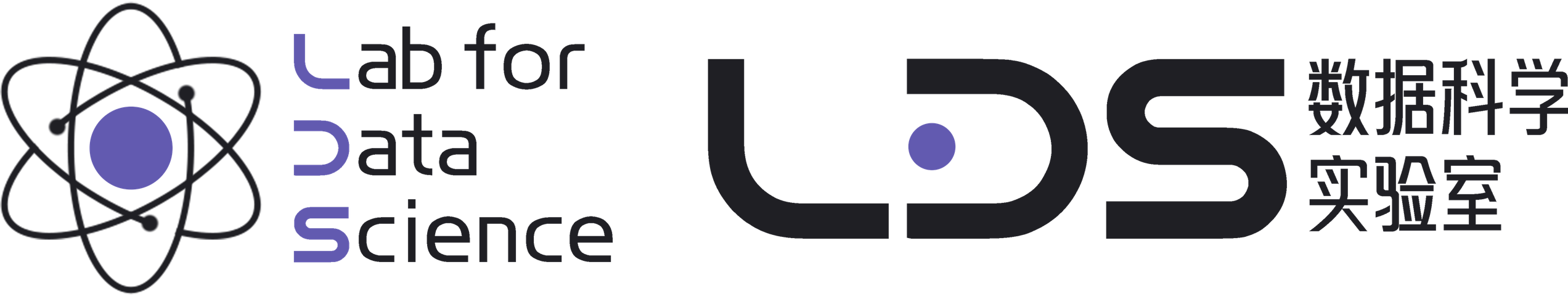}}\hspace{11mm}%
  \raisebox{-0.5\height}{\includegraphics[height=9mm]{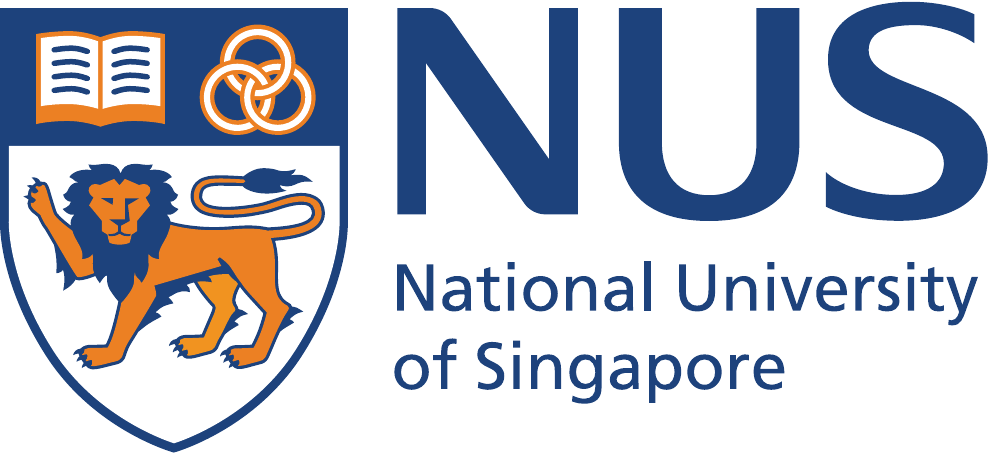}}\hspace{11mm}%
  \raisebox{-0.5\height}{\includegraphics[width=29mm]{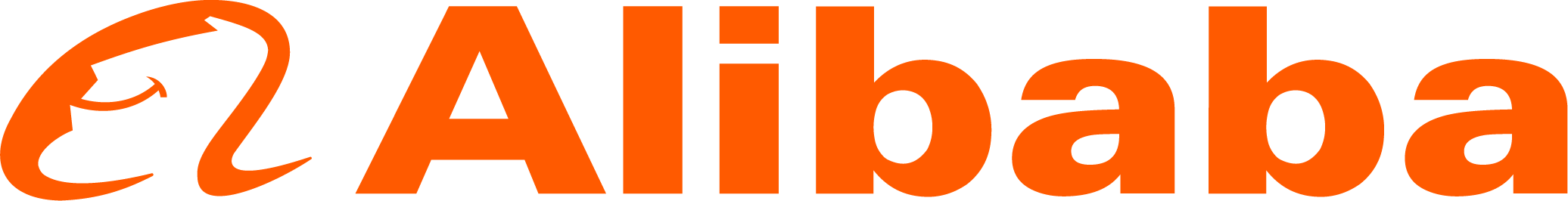}}}

\newcommand{\papertitle}{Diverse-Teacher On-Policy Distillation\\[2pt]for Generative Recommendation}
\newcommand{\paperkeywords}{Generative Recommendation, On-Policy Distillation, Popularity Bias}
\hypersetup{pdftitle={Diverse-Teacher On-Policy Distillation for Generative Recommendation},
  pdfauthor={\pdfauthors},pdfsubject={Generative recommendation},
  pdfkeywords={Generative recommendation, on-policy distillation, popularity bias}}

\begin{document}
\twocolumn[\begin{@twocolumnfalse}
  \makepreprinttitle
  \begin{preprintabstract}
    Generative recommendation generates the semantic ID (SID) of the next item from a user's interaction history. Existing training objectives mainly supervise target SID tokens, while on-policy distillation (OPD) provides denser supervision through a teacher's probability distribution over all tokens for more effective and efficient training. However, OPD can aggravate popularity bias: teacher feedback favors popular-item tokens, transferring this preference to the student, while student rollout favors popular items, leaving unpopular-item SID paths rarely generated and supervised. These two sources of bias reinforce each other and reduce recommendation diversity. We propose \methodname, a diversity-aware OPD framework. Heterogeneous teacher distillation trains two teachers specializing in popular and unpopular items and routes each training sample to the teacher matching the popularity group of its ground-truth item. The student learns from the routed teacher's probability distribution along its own generated SIDs. Diversity-aware item rollout incorporates ground-truth SIDs of unpopular items into training, covering SID paths that the student rarely generates. Only the student is used for inference. Offline experiments on an industrial dataset and three public datasets further demonstrate its effectiveness, diversity, and robustness. Moreover, we deploy \methodname in the ``Guess What You Like'' module on the 1688 homepage, where a 15-day online A/B test shows a 3.412\% increase in gross merchandise value (GMV).

  \end{preprintabstract}
  \vspace{1.4em}
\end{@twocolumnfalse}]

\section{Introduction}
\label{sec:introduction}

Generative recommendation (GR) has emerged as a promising paradigm and attracted increasing attention for its end-to-end recommendation~\cite{rajput2023recommender,wang2024letter,wang2024eager,hou2025rpg}.
Specifically, GR represents each item as a sequence of discrete tokens, \emph{i.e.,} semantic ID (SID), and directly generates the SID of the next item given a user's interaction history.
To optimize the generative recommender, the collected implicit user feedback (\emph{e.g.,} clicks) serves as the supervision signal, which trains the model to generate the item SID that matches user preference at higher probabilities.

\begin{figure}[t]
    \centering
    \makebox[\columnwidth][c]{\includegraphics[width=\columnwidth,trim=0 5 0 5.5,clip]{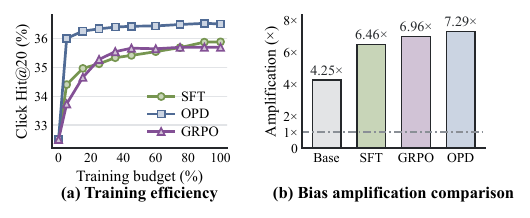}}
    \captionsetup{labelformat=empty,labelsep=none,width=\columnwidth}
    \caption[Dense feedback transfers popularity bias]{%
        \begin{minipage}{\linewidth}
        \justifying\noindent
        \leftskip=0pt\rightskip=0pt\parfillskip=0pt plus 1fil
        \spaceskip=.3333em plus .5em minus .15em\xspaceskip=\spaceskip
        \textbf{Figure~\thefigure:}
        Results on 1688. (a) OPD exhibits faster convergence and greater training data efficiency, achieving a higher click hit with a smaller training budget. (b) OPD strengthens the model’s preference for head items, leading to greater popularity bias. Amplification is the head-item generation share divided by the head-item share in the candidate set.\par
        \end{minipage}}
    \label{fig:motivation-popularity}
\end{figure}

Existing studies design different objectives to exploit implicit user feedback.
SFT maximizes the likelihood of the interacted item's SID~\cite{rajput2023recommender,wang2024letter}, DPO-based methods further contrast preferred and dispreferred items~\cite{gao2025sprec,qi2026ampo}, and RL-based methods such as GRPO reward self-generated items with user feedback~\cite{onerec2026onereason}.
Essentially, these objectives all operate on the SID tokens of target items, where each generation step is supervised by a single target token.
As a result, they over-emphasize the tokens that appear in the target SID while overlooking training signals for all other tokens.
Recently, on-policy distillation (OPD) has been explored to supervise generative recommenders with denser token-level signals~\cite{onerec2026onereason}.
Specifically, OPD lets the student model generate SIDs by itself, and employs a teacher model to provide a probability distribution over all tokens at each generation step.
In this way, tokens that appear in the SID and those that do not, will both receive training signals from the teacher, offering richer supervision for more effective and efficient model training (Figure~\ref{fig:motivation-popularity}(a)).

Despite its denser supervision, OPD suffers from the \emph{\textbf{bias aggravation}} issue in generative recommendation, \emph{i.e.,} the student model increasingly favors popular items.
As shown in Figure~\ref{fig:motivation-popularity}(b), OPD generates a larger share of popular items than SFT and GRPO.
To understand this issue, we revisit the training process of OPD.
At each optimization step, the student learns from the teacher's probability distribution, raising the generation probabilities of the tokens favored by the teacher, \emph{i.e.,} mostly the tokens of popular items~\cite{chen2023unbiased}.
Worse still, in the next rollout, the updated student generates more SIDs composed of these favored tokens, while the SIDs of less favored items are rarely generated and thus receive little supervision.
Therefore, OPD naturally inherits the teacher's bias and amplifies it round by round, resulting in poor recommendation diversity, which under-serves unpopular items and limits the exploration of users' interests.

To alleviate the bias aggravation issue, we propose a \emph{diversity-aware} OPD paradigm for generative recommendation, which follows two principles from the model and data perspectives.
\begin{itemize}[leftmargin=1.2em,labelsep=0.4em,itemindent=0pt]
    \item \textbf{Heterogeneous teachers.} Supervising all items with the probability distribution of a single teacher provides homogeneous signals, which always favor the tokens of popular items. Therefore, OPD requires heterogeneous teachers that provide different signals for different items. The key lies in leveraging the respective strengths of these teachers, so that the tokens of both popular and unpopular items receive accurate supervision.
    \item \textbf{Diverse items.} Since the student inevitably favors popular items during rollout, OPD training should cover a more diverse set of items. The key idea is to extend teacher supervision to the SID paths of unpopular items that the student rarely generates.

\end{itemize}

Following these principles, we propose a \textbf{HE}terogeneous-teacher and diverse-item \textbf{RO}llout OPD framework (\methodname) for generative recommendation, which enhances recommendation diversity from both the model and data diversity in OPD.
\emph{First}, to achieve heterogeneous teacher distillation, we train two teachers that specialize in popular and unpopular items, respectively.
During OPD, each training sample is routed to the teacher matching the popularity group of its ground-truth item, and the student learns from the probability distribution of the routed teacher over its own generated SIDs.
\emph{Second}, to balance the diversity of rollout data, we introduce a diversity-aware item rollout, which incorporates the SIDs of unpopular items into training to receive supervision from the teacher.
In this way, the student avoids overfitting the popular items favored by the teachers.
Only the student is used for inference.

Practically, we deploy \methodname in the ``Guess What You Like'' module on the 1688 homepage, where a 15-day online A/B test shows a significant 3.412\% increase in GMV (the total transaction value from both direct and indirect conversions) over the existing generative recommender.
Experiments on the 1688 industrial dataset and three public datasets further demonstrate the effectiveness, diversity, and robustness of our method.

The main contributions of this work are summarized as follows:
\begin{itemize}[leftmargin=1.2em,labelsep=0.4em,itemindent=0pt]
    \item We identify the bias aggravation issue of OPD in generative recommendation, where the student naturally inherits the teacher's bias toward popular items and amplifies it round by round.
    \item We propose a diversity-aware OPD paradigm with two principles, \emph{i.e.,} heterogeneous teachers and diverse items, and implement it as \methodname, which realizes the two principles via heterogeneous teacher distillation and diversity-aware item rollout, respectively.
    \item We deploy \methodname in the ``Guess What You Like'' module on the 1688 homepage, where an online A/B test shows a 3.412\% increase in GMV. Offline experiments on an industrial dataset and three public datasets further validate the effectiveness, diversity, and robustness of \methodname.
\end{itemize}

\section{Preliminaries}
\label{sec:preliminaries}

\subsection{Generative Recommendation}
\label{sec:pre-gr}

We consider a sequential recommendation setting where each training instance consists of a user context and a target item. Let $x=(i_1,i_2,\ldots,i_m)$ denote a history of $m$ interacted items, and let $i^+$ denote the next item to be recommended. Traditional recommendation models usually score candidate items directly. In contrast, generative recommendation formulates recommendation as a conditional generation problem, where the model generates an item identifier conditioned on the user context.

Following recent generative retrieval-based recommender systems~\cite{rajput2023recommender}, each item $i$ is represented by a semantic identifier (SID), denoted as
\begin{equation}
    s(i) = (z_1, z_2, \ldots, z_L),
    \label{eq:sid}
\end{equation}
where $z_\ell$ is the discrete token at SID position $\ell$, and $L$ is the SID length. The SID can be viewed as a tokenized representation of an item, enabling a generative model to recommend items by decoding token sequences rather than selecting from a flat item vocabulary.

Given user context $x$, a generative recommender with parameters $\theta$ defines the probability of an item by the probability of its SID:
\begin{equation}
    p_\theta(s(i)\mid x)
    = \prod_{\ell=1}^{L} p_\theta(z_\ell \mid x, z_{<\ell}),
    \label{eq:gr-factorization}
\end{equation}
where $z_{<\ell}=(z_1,\ldots,z_{\ell-1})$ denotes the prefix before position $\ell$. At inference time, the model generates an SID autoregressively, and the decoded SID is mapped back to the corresponding item.

For a target item $i^+$, we write its ground-truth SID as $y=s(i^+)=(y_1,\ldots,y_L)$ and denote the supervised training set by $\mathcal{D}_{\mathrm{SFT}}=\{(x,y)\}$. Supervised fine-tuning (SFT) minimizes the negative log-likelihood of this SID under teacher forcing:
\begin{equation}
    \mathcal{L}_{\mathrm{SFT}}(\theta)
    = - \mathbb{E}_{(x,y)\sim\mathcal{D}_{\mathrm{SFT}}}
    \sum_{\ell=1}^{L}
    \log p_\theta(y_\ell \mid x, y_{<\ell}),
    \label{eq:sft}
\end{equation}
where $y_{<\ell}=(y_1,\ldots,y_{\ell-1})$ is the ground-truth prefix. This objective provides direct supervision on the ground-truth path, but the supervision at each step is one-hot and is restricted to observed target items. Therefore, SFT is usually followed by additional training strategies that provide richer guidance to the student model.

\subsection{On-Policy Distillation in Generative Recommendation}
\label{sec:pre-opd}

On-policy distillation (OPD) further improves a student model by transferring token-level knowledge from a teacher model along trajectories generated by the student itself~\cite{agarwal2024onpolicy}. Let $p_\theta$ and $p_\phi$ denote the student's and teacher's conditional next-token distributions, respectively. For a given context $x$, the student generates an SID trajectory
\begin{equation}
    \hat{y}=(\hat{y}_1,\hat{y}_2,\ldots,\hat{y}_L) \sim p_\theta(\cdot\mid x).
    \label{eq:student-trajectory}
\end{equation}
Unlike SFT, which trains only on the ground-truth prefix $y_{<\ell}$, OPD trains the student on its own generated prefixes $\hat{y}_{<\ell}=(\hat{y}_1,\ldots,\hat{y}_{\ell-1})$. This makes the training distribution closer to the student's inference-time distribution.

At each generated prefix, the student is optimized to match the teacher's soft next-token distribution. Using $D_{\mathrm{KL}}(a\|b)$ to denote the Kullback--Leibler divergence from distribution $a$ to distribution $b$, a standard forward-KL OPD objective is
\begin{equation}
\begin{aligned}
    \mathcal{L}_{\mathrm{OPD}}(\theta)
    &= \mathbb{E}_{\substack{(x,y)\sim\mathcal{D}_{\mathrm{SFT}} \\
    \hat{y}\sim p_\theta(\cdot\mid x)}}
    \sum_{\ell=1}^{L} \\
    &\quad D_{\mathrm{KL}}\!\left(
    p_\phi(\cdot\mid x,\hat{y}_{<\ell})
    \;\|\;
    p_\theta(\cdot\mid x,\hat{y}_{<\ell})
    \right).
\end{aligned}
    \label{eq:opd}
\end{equation}
The soft teacher distribution contains information about alternative tokens beyond the single ground-truth token, providing denser supervision than SFT. Since the prefixes come from the current student, OPD also allows the student to learn from states it is likely to encounter during generation. The formulation above uses a single sampled trajectory; Section~\ref{sec:multi_teacher_opd} instead constructs a deterministic top-$J$ beam under the current student and averages the KL loss over the retained trajectories.

In generative recommendation, OPD can transfer recommendation knowledge from a stronger teacher into a compact student that is used alone at inference time. However, the effectiveness of OPD depends on two factors: the coverage of the student-induced rollout distribution and the quality of the teacher-provided supervision. In popularity-skewed recommendation data, a student trained by SFT may generate popular items more frequently, causing tail-item paths to receive fewer on-policy updates. Meanwhile, a single teacher may not provide equally effective supervision for both popular and tail items. These observations motivate us to revisit both the rollout process and the teacher design for OPD in generative recommendation.

\section{Method}
\label{sec:method}

\begin{figure*}[t]
    \centering
    \includegraphics[width=0.98\textwidth,trim=0 7.5 0 6,clip]{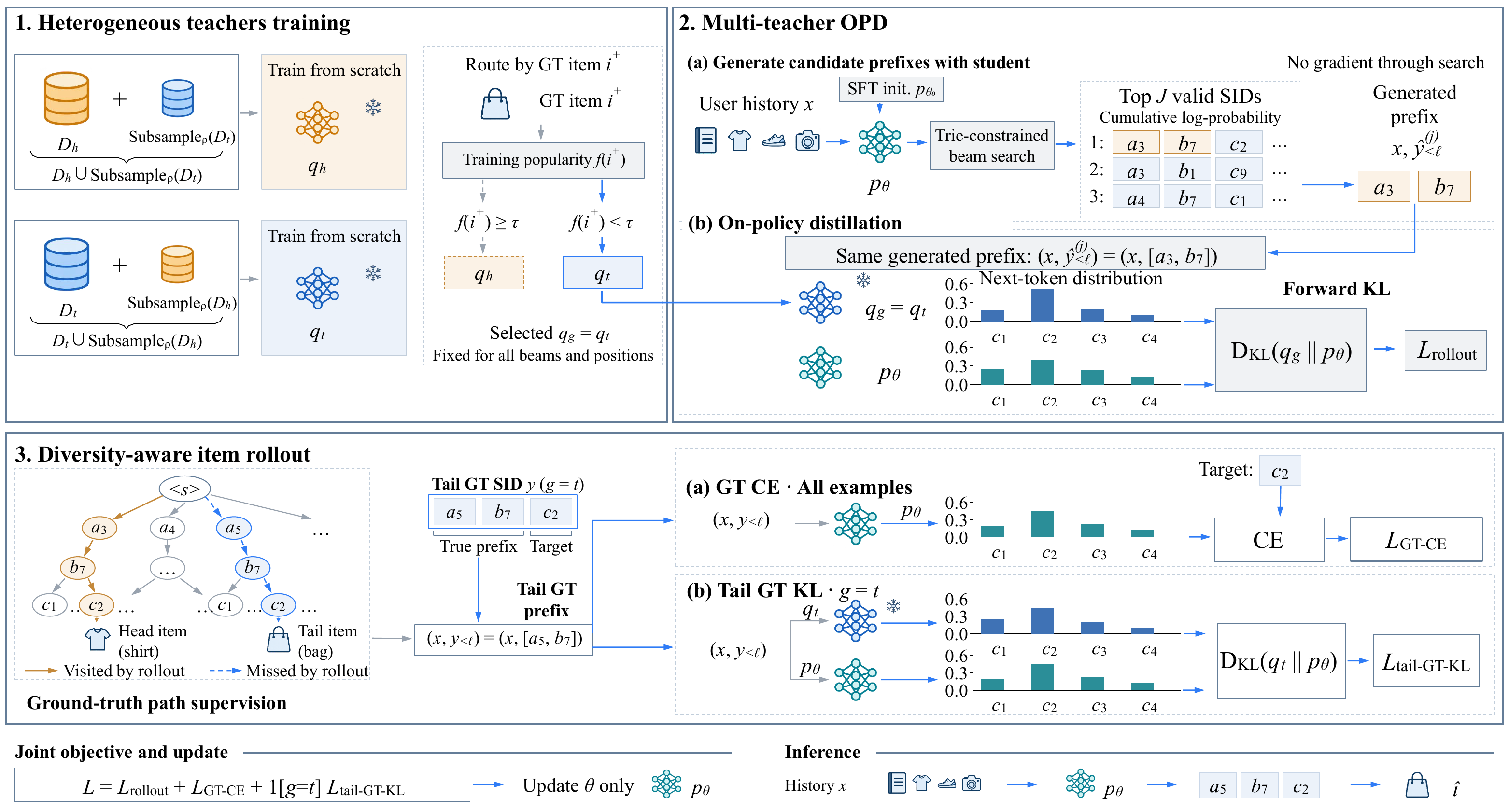}
    \caption{Overview of \methodname. (1) We train head- and tail-specialized teachers and select one according to each example's ground-truth target group. (2) The student generates SID trajectories and matches the selected teacher's distributions on generated prefixes via forward KL. (3) Ground-truth paths complement these trajectories with CE for all examples and tail-teacher KL for tail targets. All three losses update only the student, which is used alone at inference.}
    \label{fig:main-framework}
\end{figure*}

In this section, we present \methodname, a diversity-aware on-policy distillation framework for generative recommendation and it implements the two principles introduced in Section~\ref{sec:introduction} through heterogeneous-teacher distillation and diversity-aware item rollout. As illustrated in Figure~\ref{fig:main-framework}, heterogeneous teachers  provide complementary supervision for head and tail targets through specialized teachers and popularity-aware routing. Diversity-aware item rollout complements student-generated trajectories with ground-truth tail paths, bringing relevant tail items into distillation alongside the student's own generations. These two designs jointly optimize a single student model, which is used alone at inference.

\subsection{Heterogeneous-Teacher Distillation}
\label{sec:method-teacher}

Heterogeneous-teacher distillation provides supervision suited to head and tail targets through specialized teachers and popularity-aware routing, as illustrated in the first two modules of Figure~\ref{fig:main-framework}.

\subsubsection{Heterogeneous-teacher training.}

We partition $\mathcal{D}_{\mathrm{SFT}}$ into head and tail subsets $\mathcal{D}_h$ and $\mathcal{D}_t$ according to target-item frequency in the training interactions. For each example, $g\in\{h,t\}$ denotes its ground-truth target group. Let $q_h$ and $q_t$ denote the teachers' next-token distributions. We train $q_h$ on all examples in $\mathcal{D}_h$ together with a randomly sampled fraction $\rho$ of $\mathcal{D}_t$, and train $q_t$ on all of $\mathcal{D}_t$ together with the same fraction of $\mathcal{D}_h$. The cross-group fraction $\rho\in[0,1]$ retains supervision from the other group. Both teachers are trained independently from scratch using teacher-forced negative log-likelihood and remain frozen during student distillation.

\subsubsection{Multi-teacher OPD}
\label{sec:multi_teacher_opd}
For each example, routing selects $q_g=q_h$ for a head target and $q_g=q_t$ for a tail target. The selected teacher serves as the teacher distribution in Equation~\ref{eq:opd} . Given a user interaction history $x$, the current student $p_\theta$ performs trie-constrained beam search with width $J$. We retain the $J$ completed SID trajectories returned by the search, ranked by cumulative student log-probability, and denote them by $\hat{y}^{(1)},\ldots,\hat{y}^{(J)}$. On each generated prefix $\hat{y}^{(j)}_{<\ell}$, the routed teacher and the student are evaluated under the same context. As shown in the on-policy distillation branch of Figure~\ref{fig:main-framework}, their full next-token distributions are matched through forward KL divergence. The loss for the current example is
\begin{equation}
    \mathcal{L}_{\mathrm{rollout}}
    = \frac{1}{J}\sum_{j=1}^{J}\sum_{\ell=1}^{L}
    D_{\mathrm{KL}}\!\left(
    q_g(\cdot\mid x,\hat{y}^{(j)}_{<\ell})
    \;\|\;
    p_\theta(\cdot\mid x,\hat{y}^{(j)}_{<\ell})
    \right).
    \label{eq:rollout-loss}
\end{equation}
The loss sums over SID positions and averages over the retained trajectories. Generated tokens determine the conditioning prefixes, while the teacher distribution supplies the learning signal. We optimize the student distributions evaluated at these prefixes without differentiating through beam search. Throughout this section, we omit the dependence of sample losses on $(x,y)$ and $\theta$ for brevity.

\subsection{Diversity-aware Item Rollout}
\label{sec:method-rollout}

Routing adapts teacher feedback to the target group, but the coverage of on-policy distillation still depends on the paths generated by the student. A student that favors popular items may rarely reach the correct paths of tail targets, leaving relevant tail states with little distributional supervision. Diversity-aware item rollout addresses this coverage gap by complementing student-generated trajectories with ground-truth tail paths.

The tail-target example in Figure~\ref{fig:main-framework} illustrates this distinction. The student visits the generated prefix $[a_3,b_7]$, while the ground-truth tail path passes through $[a_5,b_7]$. Since routing is determined by the ground-truth target, $q_t$ supervises this example on both prefixes. The generated-path branch provides feedback on the former, and the additional ground-truth branch provides feedback on the latter.
We also retain direct supervision from the observed next item. As shown in the ground-truth CE branch of Figure~\ref{fig:main-framework}, every training example contributes the sequence-level cross-entropy loss
\begin{equation}
    \mathcal{L}_{\mathrm{GT\text{-}CE}}
    = -\sum_{\ell=1}^{L}\log p_\theta(y_\ell\mid x,y_{<\ell}).
    \label{eq:gt-ce}
\end{equation}

For every example with $g=t$, we include the ground-truth SID $y$ as an additional distillation path. Under teacher forcing, the tail teacher and the student are conditioned on the same true prefix $y_{<\ell}$, yielding
\begin{equation}
    \mathcal{L}_{\mathrm{tail\text{-}GT\text{-}KL}}
    = \sum_{\ell=1}^{L}D_{\mathrm{KL}}\!\left(
    q_t(\cdot\mid x,y_{<\ell})
    \;\|\;
    p_\theta(\cdot\mid x,y_{<\ell})
    \right).
    \label{eq:tail-gt-kl}
\end{equation}
This term supplies soft supervision along the correct tail path even when it is absent from the student's generated trajectories. Together, the two path sources cover states visited by the current student and relevant tail states supplied by the training targets, allowing both to contribute to distillation.

\subsection{Instantiation}
\label{sec:method-instantiation}

We instantiate \methodname by initializing a single student from the SFT model $p_{\theta_0}$ and jointly optimizing the three losses introduced above. The training objective is
\begin{equation}
\begin{aligned}
    \mathcal{L}(\theta)
    = \mathbb{E}_{(x,y)\sim\mathcal{D}_{\mathrm{SFT}}}
    \Bigl[{}&\mathcal{L}_{\mathrm{rollout}}
    + \mathcal{L}_{\mathrm{GT\text{-}CE}} \\
    &+ \mathbb{I}[g=t]\,
    \mathcal{L}_{\mathrm{tail\text{-}GT\text{-}KL}}\Bigr].
\end{aligned}
    \label{eq:training-objective}
\end{equation}
The three terms have unit weights, with the additional ground-truth KL applied only to tail targets. For each mini-batch, we compute the applicable losses jointly. Only the student parameters $\theta$ are updated, while both teachers remain frozen. Algorithm~\ref{alg:diverse-teacher-opd} summarizes the training procedure.

\begin{algorithm}[t]
\caption{Training \methodname}
\label{alg:diverse-teacher-opd}
\small
\renewcommand{\algorithmicrequire}{\textbf{Input:}}
\renewcommand{\algorithmicensure}{\textbf{Output:}}
\begin{algorithmic}[1]
\REQUIRE Head and tail training sets $\mathcal{D}_h,\mathcal{D}_t$,
grouped by target-item popularity; SFT student $p_{\theta_0}$;
cross-group fraction $\rho$; rollout beam width $J$
\ENSURE Distilled student $p_\theta$
\STATE \textbf{Teacher preparation:} Mix each set with a $\rho$ fraction
of the other set to form $\mathcal{D}_h^{(\rho)}$ and $\mathcal{D}_t^{(\rho)}$
\STATE Train $q_h$ and $q_t$ from scratch with teacher-forced NLL
on their respective mixed sets; freeze both teachers
\STATE Initialize the student parameters $\theta\leftarrow\theta_0$
\FOR{each mini-batch $\mathcal{B}$ from $\mathcal{D}_h\cup\mathcal{D}_t$}
    \STATE Set the batch loss $\mathcal{L}_{\mathcal{B}}\leftarrow0$
    \FOR{each example $(x,y,g)\in\mathcal{B}$}
        \STATE \textbf{Routing:} Select $q_g=q_h$ for a head target
        and $q_g=q_t$ for a tail target
        \STATE \textbf{Student rollout:} Generate $J$ trajectories
        $\widehat{\mathcal{Y}}_J$ from $p_\theta(\cdot\mid x)$
        by trie-constrained beam decoding
        \STATE \textbf{Rollout loss:} Average the prefix-wise forward KL
        $D_{\mathrm{KL}}(q_g\|p_\theta)$ over the $J$ trajectories
        \STATE \textbf{Ground-truth loss:} Sum token CE along $y$;
        for a tail target, also sum $D_{\mathrm{KL}}(q_t\|p_\theta)$
        along its ground-truth prefixes
        \STATE Add the resulting sample loss
        to $\mathcal{L}_{\mathcal{B}}$
    \ENDFOR
    \STATE \textbf{Optimization:} Update $\theta$ with one AdamW step
    on $\mathcal{L}_{\mathcal{B}}/|\mathcal{B}|$
\ENDFOR
\RETURN Distilled student $p_\theta$
\end{algorithmic}
\end{algorithm}

\section{Experiments}
\label{sec:experiments}

\setcounter{topnumber}{2}
\setcounter{bottomnumber}{1}

We conduct experiments to answer three research questions:
\begin{itemize}
    \item \textbf{RQ1:} How does \methodname compare with representative recommendation and post-training methods?
    \item \textbf{RQ2:} How does each component contribute to recommendation accuracy and popularity bias alleviation?
    \item \textbf{RQ3:} How does \methodname affect generation and accuracy across item popularity groups?
\end{itemize}

\begin{table*}[t]
\centering
\caption{Main results on 1688 and three public datasets. On 1688, Click, Pay, and Tail denote Click Hit, Pay Hit, and Tail Click Hit. Hit, Tail, and ORR are ratios (e.g., 0.52 denotes 52\%); ORR is the share of slots occupied by the five most frequent recommendations. $\mathrm{DTR}^{*}=|\mathrm{DTR}-1|$ measures the deviation of the demand-to-recommendation ratio from 1. $\uparrow$ and $\downarrow$ indicate that higher and lower values are better. Best and second-best results within each dataset are marked in bold and underlined.}
\label{tab:main-results}
\small
\setlength{\tabcolsep}{1.5pt}
\renewcommand{\arraystretch}{1.12}
\resizebox{\textwidth}{!}{%
\begin{tabular}{@{}l*{17}{c}@{}}
\toprule
& \multicolumn{5}{c}{1688}
& \multicolumn{4}{c}{Beauty}
& \multicolumn{4}{c}{Pet}
& \multicolumn{4}{c}{MovieLens-1M} \\
\cmidrule(lr){2-6}
\cmidrule(lr){7-10}
\cmidrule(lr){11-14}
\cmidrule(lr){15-18}
Method
& Click@240$\uparrow$
& Pay@240$\uparrow$
& Tail@240$\uparrow$
& $\mathrm{DTR}^{*}\downarrow$
& ORR$\downarrow$
& Hit@20$\uparrow$
& Tail@20$\uparrow$
& $\mathrm{DTR}^{*}\downarrow$
& ORR$\downarrow$
& Hit@20$\uparrow$
& Tail@20$\uparrow$
& $\mathrm{DTR}^{*}\downarrow$
& ORR$\downarrow$
& Hit@20$\uparrow$
& Tail@20$\uparrow$
& $\mathrm{DTR}^{*}\downarrow$
& ORR$\downarrow$ \\
\midrule

\rowcolor{recommendationband}
\multicolumn{18}{l}{\emph{Recommendation baselines}} \\
SASRec
& 0.5638 & 0.7426 & 0.5184 & 0.8630 & 0.0044
& 0.0212 & 0.0023 & 13.2360 & 0.0709
& 0.0240 & 0.0008 & 45.5360 & 0.0945
& 0.1762 & 0.0889 & 5.2310 & 0.0731 \\
TIGER
& 0.5598 & 0.7396 & 0.5139 & 0.7390 & 0.0040
& 0.0301 & 0.0056 & 21.9270 & 0.0490
& 0.0306 & 0.0044 & 25.2540 & 0.0781
& 0.1687 & 0.0720 & 6.9230 & 0.0858 \\
SFT
& 0.5742 & 0.7562 & 0.5302 & 0.6540 & \underline{0.0027}
& 0.0302 & 0.0061 & \underline{13.2030} & 0.0436
& 0.0355 & 0.0094 & 8.9610 & \underline{0.0430}
& 0.1640 & 0.0653 & 5.7140 & 0.0621 \\
SFT++
& 0.5723 & 0.7553 & 0.5288 & 0.6280 & 0.0028
& 0.0311 & 0.0063 & 13.6270 & 0.0494
& 0.0348 & 0.0085 & 8.6130 & 0.0441
& 0.1693 & 0.0681 & 6.1310 & 0.0638 \\
\addlinespace[1pt]

\rowcolor{distillationband}
\multicolumn{18}{l}{\emph{Distillation baselines}} \\
GT-KD
& 0.5766 & 0.7562 & 0.5230 & 0.9940 & 0.0030
& \underline{0.0373} & 0.0072 & 22.4950 & 0.0384
& 0.0357 & 0.0109 & 5.2730 & 0.0540
& 0.1758 & 0.0734 & 4.3910 & 0.0516 \\
Off-policy KD
& 0.5759 & 0.7563 & 0.5294 & 0.7820 & 0.0030
& 0.0371 & 0.0089 & 16.4800 & 0.0395
& 0.0356 & 0.0110 & \underline{3.7740} & 0.0705
& 0.1792 & 0.0802 & 3.6220 & 0.0490 \\
GKD
& \underline{0.5786} & \underline{0.7584} & 0.5301 & 0.7150 & 0.0029
& 0.0369 & 0.0086 & 16.0960 & \underline{0.0366}
& 0.0355 & 0.0092 & 5.3270 & 0.0733
& 0.1771 & 0.0811 & 3.6360 & 0.0487 \\
MOPD
& 0.5731 & 0.7558 & 0.5230 & 0.7950 & 0.0029
& 0.0341 & 0.0069 & 18.3680 & 0.0474
& 0.0361 & 0.0103 & 4.3290 & 0.0692
& 0.1765 & 0.0785 & 3.8360 & 0.0494 \\
\addlinespace[1pt]

\rowcolor{debiasband}
\multicolumn{18}{l}{\emph{Preference optimization, debiasing, and search baselines}} \\
SID-GRPO
& 0.5692 & 0.7408 & 0.4998 & 0.7560 & 0.0035
& 0.0329 & \underline{0.0108} & 15.7610 & 0.0399
& 0.0348 & 0.0088 & 4.2800 & 0.0581
& 0.1540 & 0.0789 & \underline{2.8830} & 0.0563 \\
Ghost-AUO
& 0.5778 & 0.7577 & \underline{0.5355} & 0.5550 & 0.0028
& \underline{0.0373} & 0.0089 & 15.4290 & 0.0508
& 0.0364 & \underline{0.0121} & 5.7200 & 0.0516
& 0.1775 & 0.0827 & 3.5430 & 0.0479 \\
BEAR
& 0.5763 & 0.7571 & 0.5351 & \underline{0.4790} & 0.0028
& 0.0372 & 0.0095 & 15.2870 & 0.0510
& \underline{0.0370} & 0.0115 & 8.3230 & 0.0546
& \underline{0.1896} & \underline{0.0954} & 3.0800 & \textbf{0.0418} \\
\midrule

\rowcolor{oursgray}
\textbf{\methodname}
& \textbf{0.5819} & \textbf{0.7607} & \textbf{0.5416} & \textbf{0.3920} & \textbf{0.0025}
& \textbf{0.0391} & \textbf{0.0127} & \textbf{12.4490} & \textbf{0.0365}
& \textbf{0.0384} & \textbf{0.0144} & \textbf{3.2370} & \textbf{0.0319}
& \textbf{0.1936} & \textbf{0.0995} & \textbf{2.5260} & \underline{0.0434} \\
\bottomrule
\end{tabular}%
}
\end{table*}

\subsection{Experimental Setup}
\label{sec:exp-settings}

\subsubsection{Datasets}

We evaluate HERO on 1688 and three public datasets: Beauty and Personal Care (Beauty) and Pet Supplies (Pet) from Amazon Reviews 2023~\cite{hou2024bridging}\footnote{\url{https://amazon-reviews-2023.github.io/}}, and MovieLens-1M~\cite{harper2015movielens}\footnote{\url{https://grouplens.org/datasets/movielens/1m/}}. On 1688, a large Chinese business-to-business (B2B) wholesale platform, each page view (PV) defines a recommendation instance with potentially multiple clicked and purchased targets. SFT uses the full July 16--August 4, 2026 training window, while continued SFT and distillation post-training use its final four days. Evaluation uses August 6, 2026 data. For Beauty and Pet, we retain valid ratings from September 2021 to September 2023 and remove duplicate user--item--timestamp events. All MovieLens-1M ratings are treated as implicit interactions. Across the public datasets, we iteratively apply 5-core filtering to users and items and use the most recent 50 items before each target for next-item prediction. Targets are split globally in chronological order into 80\%/10\%/10\% for training, validation, and testing, with checkpoints selected on validation data and results reported on the test split. After preprocessing, Beauty contains 80,052 users, 54,152 items, and 940,418 interactions, Pet contains 61,653 users, 27,640 items, and 523,450 interactions, and MovieLens-1M contains 6,040 users, 3,416 items, and 999,611 interactions. Head items have at least 4,096 training-period interactions on 1688 or at least 33, 44, and 783 training interactions on Beauty, Pet, and MovieLens-1M, respectively. All remaining items belong to the tail group.

\subsubsection{Evaluation Metrics}

Let $\mathcal{U}$ denote the evaluation instances and let $R_u^K=(\hat{i}_{u,1},\ldots,\hat{i}_{u,K})$ be the top-$K$ recommendation list for instance $u\in\mathcal{U}$, where $K$ is the evaluation cutoff.

\paragraph{\normalfont\bfseries Hit@$K$.}
On 1688, Click Hit@240 and Pay Hit@240 are the fractions of eligible page views (PVs) for which $R_u^{240}$ contains at least one click or payment target, respectively. Each public-data instance has one next-item target, and Hit@20 measures whether it appears in $R_u^{20}$. Tail Hit@$K$ restricts the metric to tail targets. On the public datasets, Tail Hit@20 includes tail items observed in the training set and excludes train-unseen targets.

\paragraph{\normalfont\bfseries Demand-to-Recommendation Ratio (DTR)}
For popularity group $g\in\{h,t\}$, let $\mathcal{I}_g$ be its item set and let $Y_u$ be the ground-truth demand set for instance $u$: clicked targets on 1688 and the singleton next-item target on each public dataset. With rank discount $w_r=1/\log_2(r+1)$, the recommendation-exposure share $E_g$ and demand share $M_g$ are
\begin{equation}
\begin{aligned}
E_g &= \frac{\sum_{u\in\mathcal{U}}\sum_{r=1}^{K}w_r\,\mathbb{I}[\hat{i}_{u,r}\in\mathcal{I}_g]}
{\sum_{u\in\mathcal{U}}\sum_{r=1}^{K}w_r}, \\
M_g &= \frac{\sum_{u\in\mathcal{U}}|Y_u\cap\mathcal{I}_g|}
{\sum_{u\in\mathcal{U}}|Y_u|}.
\end{aligned}
\label{eq:dtr-shares}
\end{equation}
We define $\mathrm{DTR}=(E_h/M_h)/(E_t/M_t)$. DTR$=1$ indicates exposure proportional to demand, and models are ranked by distance to 1.

\paragraph{\normalfont\bfseries ORR-Top5.}
We use ORR-Top5 to denote the fraction of all recommendation slots occupied by the five most frequently generated items across the evaluation set. A lower value indicates less concentration on a small set of repeated items.

\subsubsection{Baselines}

\textbf{SASRec}~\cite{kang2018sasrec} is a self-attentive sequential recommender that directly ranks item IDs. \textbf{TIGER}~\cite{rajput2023recommender} is a generative retriever trained with our fixed SIDs. \textbf{SFT} is our supervised generative model before post-training. \textbf{SFT++} continues supervised training from SFT on the final 20\% of public training data or the final four days on 1688, providing a comparison with continued supervised optimization.
Distillation baselines differ in their supervision paths and teacher selection. \textbf{Ground-truth-path knowledge distillation (GT-KD)} matches the routed teacher along ground-truth SID prefixes, whereas \textbf{off-policy KD} distills along trajectories from a frozen SFT behavior model. \textbf{Generalized Knowledge Distillation (GKD)}~\cite{agarwal2024onpolicy} distills along student trajectories. \textbf{Multi-Teacher On-Policy Distillation (MOPD)}~\cite{ma2026mopd} routes examples to multiple teachers.
We also compare with \textbf{SID-GRPO}~\cite{MiniOneRec}, which applies GRPO to SID sequences. \textbf{Ghost-AUO}~\cite{yin2026ghost} is a generative popularity-debiasing method, and \textbf{BEAR}~\cite{yang2026bear} is a search-aware generative training method. Detailed settings are in Appendix~\ref{app:baseline}.

\subsubsection{Model Structure}
For 1688, we use a six-layer BART-derived decoder-only prefix language model (PrefixLM) with hidden size 512, four attention heads, and 12 feed-forward experts per layer with top-3 routing. For the public datasets, we use a six-layer PrefixLM with hidden size 384, six attention heads, and eight feed-forward experts per layer with top-2 routing. Within each setting, the student and both teachers share the same architecture.

\subsubsection{Implementation Details}

Following prior work, we use three levels of semantic codes on the public datasets, while adopting four levels on 1688 to accommodate its substantially larger item catalog.
On 1688, SIDs use four position-specific codebooks of size 512. For the public datasets, we encode item text with Qwen3-Embedding-\allowbreak 8B and obtain three residual-quantization codes from 256-entry codebooks, followed by a collision token. On 1688, SFT starts from the production model and runs for one pass over the full training window using AdamW, a learning rate of $10^{-5}$, bfloat16 precision, and a batch size of 512 per GPU on eight H20. Public-data SFT is trained for five epochs. The two teachers are trained from scratch for one pass on 1688 and five epochs on the public datasets, with a cross-group fraction $\rho=10\%$. Teacher and public-data SFT checkpoints are selected by the lowest validation token-level cross-entropy. SFT++ continues from the corresponding SFT checkpoint on the final four days of 1688 training data or the final 20\% of the public training sequence. Inference uses deterministic Beam-240 on 1688 and trie-constrained Beam-20 on the public datasets.

\subsection{Overall Performance (RQ1)}
\label{sec:exp-overall}

Table~\ref{tab:main-results} reports the offline results on 1688 and three public datasets. \methodname consistently achieves the highest overall and tail accuracy across all four datasets, demonstrating its effectiveness across different recommendation scenarios. These improvements are consistent with the two considerations in our design: providing teacher feedback tailored to the target item's popularity group and extending supervision to relevant tail-item paths that the student rarely generates.

Improvements in overall accuracy do not necessarily translate into better tail recommendation or lower popularity bias. On 1688, GKD improves overall click accuracy but leaves tail accuracy largely unchanged, while increasing exposure imbalance and output concentration. Continued SFT on the same recent data also reduces all three accuracy metrics. These comparisons show that additional supervised training or standard distillation does not necessarily improve recommendation accuracy and exposure balance jointly.

Furthermore, \methodname achieves the DTR closest to 1 on all four datasets and consistently reduces ORR-Top5. Notably, compared with standard OPD (GKD), \methodname also lowers both DTR$^*$ and ORR-Top5 on all four datasets.  Together with the accuracy results, this demonstrates that better overall and tail recommendation can be achieved alongside more balanced exposure and less concentrated outputs. On MovieLens-1M, BEAR achieves the lowest ORR-Top5, while \methodname provides better overall and tail accuracy and better demand--exposure alignment, indicating that lower output concentration alone does not imply better recommendation quality.

\subsection{Component Contributions to Accuracy and Bias Alleviation (RQ2)}
\label{sec:exp-components}

\subsubsection{Component Ablation}

\begin{table}[t]
\centering
\caption{Component ablation on 1688. Metrics follow Table~\ref{tab:main-results}.}
\label{tab:ablation}
\scriptsize
\setlength{\tabcolsep}{4pt}
\resizebox{\columnwidth}{!}{%
\begin{tabular}{l*{5}{c}}
\toprule
Variant & Click@240$\uparrow$ & Pay@240$\uparrow$ & Tail@240$\uparrow$ & $|\mathrm{DTR}-1|\downarrow$ & ORR$\downarrow$ \\
\midrule
\rowcolor{oursgray}
\textbf{Ours} & \textbf{0.5819} & \textbf{0.7607} & \textbf{0.5416} & \textbf{0.3920} & \textbf{0.0025} \\
w/o rollout KL & 0.5763 & 0.7562 & 0.5222 & 0.4280 & 0.0027 \\
Head rollout teacher & 0.5715 & 0.7496 & 0.4977 & 2.7360 & 0.0037 \\
Tail rollout teacher & 0.5463 & 0.7213 & 0.5344 & 0.4780 & 0.0027 \\
50/50 rollout mixture & 0.5778 & 0.7571 & 0.5252 & 1.2280 & 0.0031 \\
w/o GT-CE & 0.5753 & 0.7551 & 0.5387 & 0.4630 & 0.0027 \\
w/o tail-path GT-KL & 0.5751 & 0.7570 & 0.5329 & 0.7420 & 0.0027 \\
\bottomrule
\end{tabular}}
\end{table}

We compare three alternative teacher configurations for rollout KL, all retaining GT-CE: using $q_h$ for every example, using $q_t$ for every example, and using the fixed mixture $\frac{1}{2}q_h+\frac{1}{2}q_t$. The head-teacher variant also omits tail-path GT-KL. The full model selects the teacher according to the target item's popularity group.

As shown in Table~\ref{tab:ablation}, the single-teacher configurations exhibit different accuracy--bias trade-offs, while target-aware routing achieves better overall and tail accuracy with more balanced exposure. Moreover, routing outperforms the fixed teacher mixture, suggesting that matching teacher feedback to the target group is important for effectively utilizing heterogeneous teachers.

The loss ablations show that removing any of the three supervision signals degrades performance, supporting their complementary contributions. In particular, removing rollout KL causes the largest tail-accuracy drop among the loss ablations on 1688, highlighting the importance of teacher guidance on student-generated trajectories. Removing tail-path GT-KL also weakens tail retrieval and exposure balance, indicating the benefit of extending soft supervision to correct tail paths that student rollout may miss. Further analysis of tail-prefix coverage is presented in Figure~\ref{fig:tail-prefix-coverage}, and public-dataset ablations are provided in Appendix~\ref{app:public-ablation}.

\begin{figure}[t]
    \centering
    \includegraphics[width=\linewidth,trim=0 6 0 10,clip]{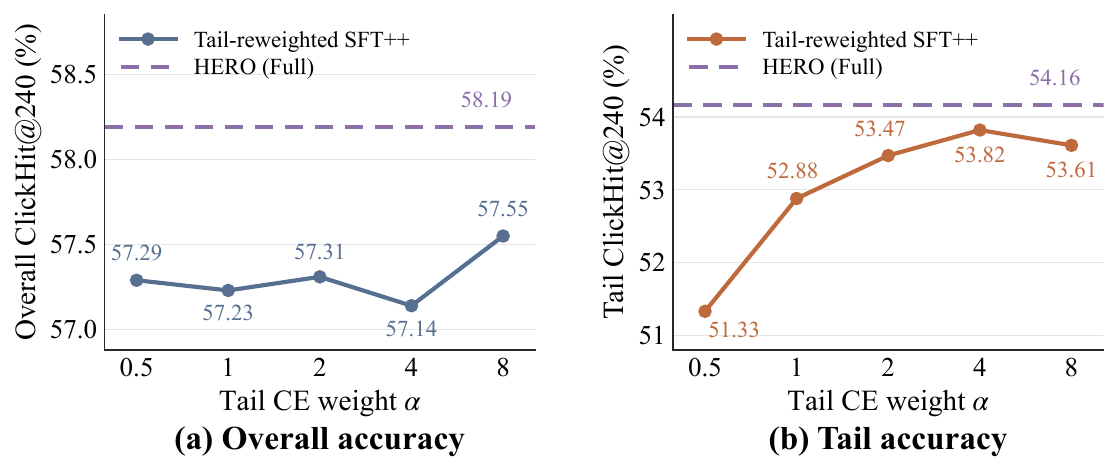}
    \caption{Effect of tail-target CE reweighting on 1688.
    (a) Overall ClickHit@240.
    (b) Tail ClickHit@240.
    Dashed lines indicate the performance of HERO.}
    \label{fig:tail_ce}
\end{figure}

\subsubsection{Comparison with Tail-reweighted SFT}
To examine whether simply adjusting the loss weight for tail items can match the performance of HERO, we retain the SFT++ training setup and vary the tail-target CE weight over \(\{0.5,1,2,4,8\}\), while fixing the head-target weight to 1. As shown in Figure~\ref{fig:tail_ce}, simply reweighting the loss in SFT fails to match the performance of HERO, showing the effectiveness of HERO to both improve overall hit and tail hit.

\subsubsection{Head--Tail Accuracy Trade-off}

We compare single-teacher OPD with target-aware routing, with and without tail-path GT-KL, on MovieLens-1M.

\begin{figure}[htbp]
    \centering
    \includegraphics[width=\columnwidth,trim=0 5 0 5.5,clip]{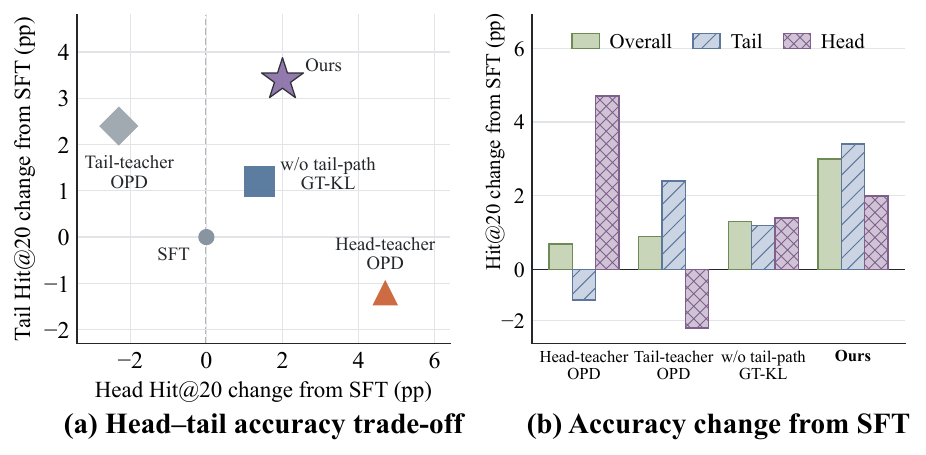}
    \caption{Changes in overall, head, and tail Hit@20 relative to SFT on MovieLens-1M.}
    \label{fig:bias-mitigation}
\end{figure}

As shown in Figure~\ref{fig:bias-mitigation}, the single-teacher variants improve accuracy on one popularity group at the expense of the other. Target-aware routing improves both head and tail accuracy over SFT, and incorporating tail-path GT-KL further improves both groups. These observations support the complementary roles of teacher routing and tail-path supervision in alleviating the head--tail trade-off. In particular, additional supervision for tail items strengthens tail retrieval while also benefiting head recommendation.

\subsection{Mechanism Analysis (RQ3)}
\label{sec:exp-analysis}

We next examine how teacher feedback affects the popularity composition of student generations and whether tail-path supervision improves coverage of correct tail prefixes. We then analyze relative accuracy gains across item popularity strata.

\subsubsection{Distributional Feedback Density}

\begin{figure}[htbp]
    \centering
    \includegraphics[width=\columnwidth,trim=0 5 0 5.5,clip]{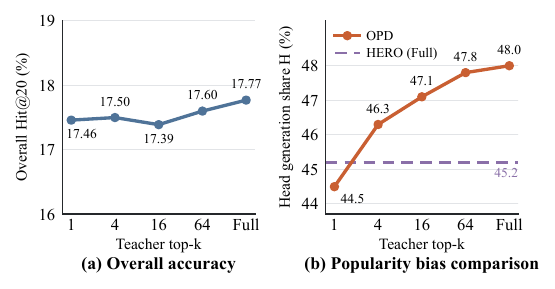}
    \caption{Effect of retaining the teacher's top-$k$ candidates on MovieLens-1M. (a) Overall Hit@20 varies only slightly. (b) Head-item generation share increases with $k$.}
    \label{fig:feedback-density}
\end{figure}

To examine how feedback density affects the transfer of a teacher's popularity preference, we use a teacher trained on MovieLens-1M. We retain all SID prediction positions and vary the number of explicit teacher candidates from top-1 to the full vocabulary. Figure~\ref{fig:feedback-density} compares overall accuracy and the head-item generation share across these settings.

\begin{samepage}
As shown in Figure~\ref{fig:feedback-density}, increasing the feedback density of the teacher shifts student generations toward head items, while overall accuracy changes little. This suggests that richer distributional feedback can reinforce the teacher's popularity preference in student outputs. Such a shift may be overlooked when distillation is evaluated solely by overall accuracy, highlighting the importance of examining the popularity distribution of generated items. In contrast, with the same full-vocabulary feedback, \methodname{} reduces the head-item generation share, mitigating this shift while retaining dense supervision.
\par\end{samepage}

\subsubsection{Tail-Prefix Reachability under Free Rollout}

On MovieLens-1M, we examine whether the improved tail accuracy is accompanied by better coverage of correct tail prefixes during free student generation. We compare SFT, Ours without tail-path GT-KL, and Ours under free Beam-32 rollout. After generating $\ell$ SID tokens, prefix coverage is the fraction of tail-target contexts for which at least one trajectory matches the ground-truth prefix $y_{\leq\ell}$.

\begin{figure}[htbp]
    \centering
    \includegraphics[width=0.95\columnwidth,trim=0 5 0 5,clip]{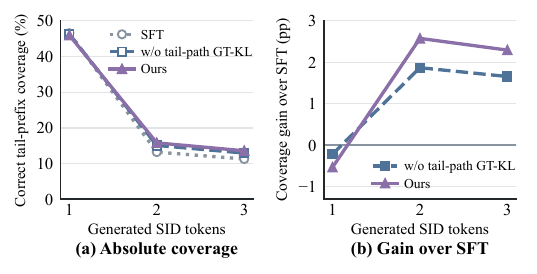}
    \caption{Correct tail-prefix coverage under free Beam-32 rollout. (a) Absolute coverage after each generated SID token. (b) Coverage gain over SFT.}
    \label{fig:tail-prefix-coverage}
\end{figure}

As shown in Figure~\ref{fig:tail-prefix-coverage}, the models exhibit similar coverage at the first SID position, while \methodname achieves higher coverage of correct tail prefixes at later positions. Adding tail-path GT-KL further improves coverage over the variant without this loss, supporting the effectiveness of additional tail-path supervision. These observations suggest that supervision on ground-truth tail paths helps the student retain correct tail continuations as decoding proceeds, translating into better coverage during free rollout.

\subsubsection{Accuracy Gains across Item Popularity Strata}

\begin{figure}[htbp]
    \centering
    \includegraphics[width=\columnwidth,trim=0 5 0 5.5,clip]{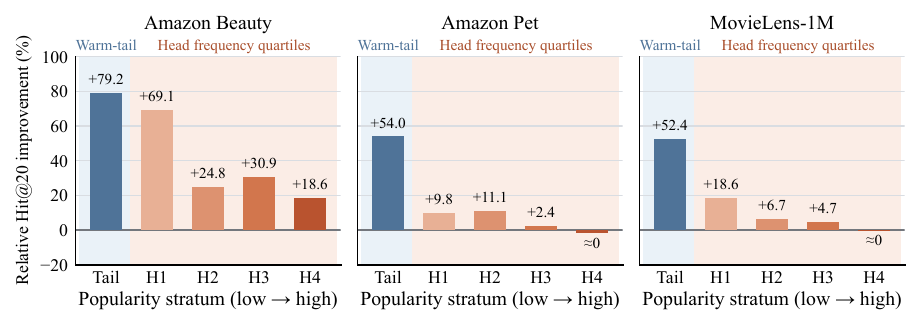}
    \caption{Relative Hit@20 improvement over SFT by target-item popularity stratum. Tail contains warm-tail targets with at least two training interactions; H1--H4 are head-item frequency quartiles from low to high.}
    \label{fig:item-popularity-gain}
\end{figure}

To identify where accuracy improves across popularity levels, we stratify public-dataset targets by training frequency and measure the relative Hit@20 improvement over SFT. We exclude targets with fewer than two training interactions, retain the remaining tail targets as the warm-tail stratum, and divide head targets into four frequency quartiles. Figure~\ref{fig:item-popularity-gain} reports the relative gains in each stratum.

As shown in Figure~\ref{fig:item-popularity-gain}, tail targets receive the largest relative improvements across all three datasets, while gains generally diminish as head-item popularity increases. These results support the effectiveness of \methodname in improving recommendation for less frequent items. Performance on the highest-frequency groups in Pet and MovieLens-1M remains nearly unchanged, indicating that the benefits are more pronounced for less popular items. Appendix~\ref{app:user-activity} provides a complementary analysis by user activity.

\subsection{Online A/B Test}
\label{sec:exp-online}

We deployed \methodname in the ``Guess What You Like'' recommendation module on the 1688 homepage and conducted an online A/B test from September 24 to October 9, 2026. We report results for three relevant arms: \methodname, a traditional non-generative recommender, and an existing generative recommender that does not use \methodname. Assignment was randomized at the user level. The three arms received equal traffic over 15 days.

Compared with the existing generative recommender before further training with our method, \methodname increases GMV by 3.412\% ($p=0.0104$), with a 95\% confidence interval of $[0.759\%, 6.065\%]$. The comparison with the traditional recommender provides a broader system-level reference: GMV increases by 25.3\%, with a 95\% confidence interval of $[21.63\%, 28.982\%]$ and $p=1.84\times10^{-50}$. Online inference uses only the distilled student and introduces no material change in response time relative to the generative control.

\begin{figure}[htbp]
    \centering
    \includegraphics[width=0.85\linewidth]{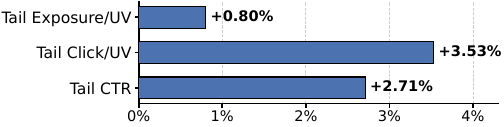}
    \caption{Relative improvement of \methodname over the online baseline on tail-item metrics. The online baseline is the existing generative recommender without \methodname.}
    \label{fig:online-tail}
\end{figure}

We also evaluate online tail-item performance using three metrics: Tail Exposure/UV (exposures per user), Tail Click/UV (clicks per user), and Tail CTR (clicks per exposure). As shown in Figure~\ref{fig:online-tail}, the model trained with \methodname improves all three metrics over the online baseline, increasing tail-item exposure and engagement while improving click efficiency.

\section{Related Work}
\label{sec:related-work}

\subsection{On-Policy Distillation for Generative Recommendation}
\label{sec:rw-opd}

Generative recommendation generates item identifiers from user histories using semantic, textual, or multi-facet representations~\cite{rajput2023recommender,tan2024idgenrec,lin2024transrec}. Existing studies improve recommendation representations, item tokenization, and generation through semantic--collaborative modeling and order-agnostic or parallel decoding~\cite{liao2024llara,wang2024letter,wang2024eager,lin2025setrec,hou2025rpg}. These models typically rely on teacher forcing with one-hot supervision. To enrich supervision, conventional methods transfer rankings or soft predictions on fixed user--item examples~\cite{tang2018ranking,lee2019collaborative,lee2021dual}, with extensions to continual and multimodal settings~\cite{lee2024continual,zhuang2025wasserstein}. Preference optimization instead learns from comparisons between complete recommendations~\cite{qi2026ampo}. These sequence-level objectives distinguish preferred from dispreferred outputs, but do not directly provide a teacher distribution at each generation step.

On-policy distillation (OPD) provides richer token-level supervision by querying a teacher along trajectories generated by the current student~\cite{agarwal2024onpolicy}. MOPD extends this approach to multiple teachers specialized by domain and routes each prompt to its corresponding expert~\cite{ma2026mopd}. These approaches emphasize distributional feedback and teacher capability coverage. For generative recommendation, however, teacher feedback may favor popular items, while student rollout may rarely visit tail-item paths. Our work focuses on these two limitations, examining how the source and coverage of on-policy feedback affect popularity bias.

\subsection{Debiased Recommendation}
\label{sec:rw-debias}

Popularity bias can lead to disproportionate exposure of popular items. Prior studies examine its relation to calibration, echo chambers, and temporal dynamics~\cite{abdollahpouri2020connection,ge2020echo,zhu2021dynamic}. General approaches address biased or noisy feedback through propensity correction and denoising~\cite{saito2020unbiased,wang2021denoising}. Popularity debiasing methods further separate interest from conformity or adjust popularity effects through causal modeling~\cite{zheng2021dice,zhang2021pda,wang2021deconfounded,wei2021macr}, while other approaches use representation learning, debiasing objectives, or spectral regularization~\cite{xv2022neutralizing,zhang2023popularity,lin2025spectral}.

Distillation and autoregressive generation introduce additional channels for bias transfer. Counterfactual recommendation transfers supervision from uniformly collected data~\cite{liu2020counterfactual}. UnKD shows that biased teacher soft labels can propagate and intensify popularity bias and transfers ranking knowledge within popularity strata to mitigate it~\cite{chen2023unbiased}. For generative recommendation, D$^3$ modifies decoding to reduce amplification and homogeneous outputs~\cite{bao2024decoding}, while SPRec uses self-play preference optimization to reduce biased recommendations~\cite{gao2025sprec}. \methodname focuses on bias in token-level feedback along student-generated trajectories. It selects teachers according to target popularity and supplements rollout with ground-truth tail paths, addressing the source and coverage of distillation feedback.

\section{Conclusion}
In this work, we identified how on-policy distillation can amplify popularity bias in generative recommendation through biased teacher feedback and insufficient coverage of tail-item paths. To address these two sources, we proposed HERO, which combines popularity-specialized teachers and target-aware routing with auxiliary distillation on ground-truth tail paths. Experiments on 1688 and three public datasets show improvements in overall and tail recommendation accuracy, alongside better alignment of recommendation exposure with observed demand across popularity groups. Ablation and mechanism analyses support the complementary roles of teacher routing and tail-path supervision. A 15-day online A/B test on 1688 further demonstrates a 3.412\%
increase in GMV over the existing generative recommender, with no material change in response time. These results highlight the importance of jointly improving teacher feedback and supervision coverage for effective on-policy distillation in generative recommendation.

\label{sec:conclusion}

\bibliographystyle{plainnat}
\bibliography{references}

\begin{thebibliography}{39}
\providecommand{\natexlab}[1]{#1}
\providecommand{\url}[1]{\texttt{#1}}
\expandafter\ifx\csname urlstyle\endcsname\relax
  \providecommand{\doi}[1]{doi: #1}\else
  \providecommand{\doi}{doi: \begingroup \urlstyle{rm}\Url}\fi

\bibitem[Abdollahpouri et~al.(2020)Abdollahpouri, Mansoury, Burke, and
  Mobasher]{abdollahpouri2020connection}
Himan Abdollahpouri, Masoud Mansoury, Robin Burke, and Bamshad Mobasher.
\newblock The connection between popularity bias, calibration, and fairness in
  recommendation.
\newblock In \emph{Proceedings of the 14th ACM Conference on Recommender
  Systems}, pages 726--731. Association for Computing Machinery, 2020.
\newblock \doi{10.1145/3383313.3418487}.
\newblock URL \url{https://doi.org/10.1145/3383313.3418487}.

\bibitem[Agarwal et~al.(2024)Agarwal, Vieillard, Zhou, Stanczyk, Ramos~Garea,
  Geist, and Bachem]{agarwal2024onpolicy}
Rishabh Agarwal, Nino Vieillard, Yongchao Zhou, Piotr Stanczyk, Sabela
  Ramos~Garea, Matthieu Geist, and Olivier Bachem.
\newblock On-policy distillation of language models: Learning from
  self-generated mistakes.
\newblock In \emph{International Conference on Learning Representations}, pages
  21246--21263, 2024.
\newblock URL
  \url{https://proceedings.iclr.cc/paper_files/paper/2024/hash/5be69a584901a26c521c2b51e40a4c20-Abstract-Conference.html}.

\bibitem[Bao et~al.(2024)Bao, Zhang, Zhang, Huo, Chen, and
  Feng]{bao2024decoding}
Keqin Bao, Jizhi Zhang, Yang Zhang, Xinyue Huo, Chong Chen, and Fuli Feng.
\newblock Decoding matters: Addressing amplification bias and homogeneity issue
  in recommendations for large language models.
\newblock In \emph{Proceedings of the 2024 Conference on Empirical Methods in
  Natural Language Processing}, pages 10540--10552. Association for
  Computational Linguistics, 2024.
\newblock \doi{10.18653/v1/2024.emnlp-main.589}.
\newblock URL \url{https://doi.org/10.18653/v1/2024.emnlp-main.589}.

\bibitem[Chen et~al.(2023)Chen, Chen, Feng, Zhou, and He]{chen2023unbiased}
Gang Chen, Jiawei Chen, Fuli Feng, Sheng Zhou, and Xiangnan He.
\newblock Unbiased knowledge distillation for recommendation.
\newblock In \emph{Proceedings of the Sixteenth {ACM} International Conference
  on Web Search and Data Mining}, pages 976--984. Association for Computing
  Machinery, 2023.
\newblock \doi{10.1145/3539597.3570477}.
\newblock URL \url{https://doi.org/10.1145/3539597.3570477}.

\bibitem[Gao et~al.(2025)Gao, Chen, Yuan, Huang, Yu, and He]{gao2025sprec}
Chongming Gao, Ruijun Chen, Shuai Yuan, Kexin Huang, Yuanqing Yu, and Xiangnan
  He.
\newblock {SPRec}: Self-play to debias {LLM}-based recommendation.
\newblock In \emph{Proceedings of the {ACM} Web Conference 2025}, pages
  5075--5084. Association for Computing Machinery, 2025.
\newblock \doi{10.1145/3696410.3714524}.
\newblock URL \url{https://doi.org/10.1145/3696410.3714524}.

\bibitem[Ge et~al.(2020)Ge, Zhao, Zhou, Pei, Sun, Ou, and Zhang]{ge2020echo}
Yingqiang Ge, Shuya Zhao, Honglu Zhou, Changhua Pei, Fei Sun, Wenwu Ou, and
  Yongfeng Zhang.
\newblock Understanding echo chambers in e-commerce recommender systems.
\newblock In \emph{Proceedings of the 43rd International ACM SIGIR Conference
  on Research and Development in Information Retrieval}, pages 2261--2270.
  Association for Computing Machinery, 2020.
\newblock \doi{10.1145/3397271.3401431}.
\newblock URL \url{https://doi.org/10.1145/3397271.3401431}.

\bibitem[Harper and Konstan(2015)]{harper2015movielens}
F.~Maxwell Harper and Joseph~A. Konstan.
\newblock The {MovieLens} datasets: History and context.
\newblock \emph{ACM Transactions on Interactive Intelligent Systems},
  5\penalty0 (4), 2015.
\newblock \doi{10.1145/2827872}.
\newblock URL \url{https://doi.org/10.1145/2827872}.

\bibitem[Hou et~al.(2025)Hou, Li, Shin, Jeon, Santhanam, Shao, Hassani, Yao,
  and McAuley]{hou2025rpg}
Yupeng Hou, Jiacheng Li, Ashley Shin, Jinsung Jeon, Abhishek Santhanam, Wei
  Shao, Kaveh Hassani, Ning Yao, and Julian McAuley.
\newblock Generating long semantic ids in parallel for recommendation.
\newblock In \emph{Proceedings of the 31st ACM SIGKDD Conference on Knowledge
  Discovery and Data Mining}, pages 956--966. Association for Computing
  Machinery, 2025.
\newblock \doi{10.1145/3711896.3736979}.
\newblock URL \url{https://doi.org/10.1145/3711896.3736979}.

\bibitem[Hou et~al.(2026)Hou, Li, Fu, He, Yan, Chen, and
  McAuley]{hou2024bridging}
Yupeng Hou, Jiacheng Li, Xiangjun Fu, Zhankui He, An~Yan, Xiusi Chen, and
  Julian McAuley.
\newblock Bridging language and items for retrieval and recommendation:
  Benchmarking {LLMs} as semantic encoders.
\newblock In \emph{Proceedings of the 64th Annual Meeting of the Association
  for Computational Linguistics (Volume 1: Long Papers)}, pages 3251--3265.
  Association for Computational Linguistics, 2026.
\newblock \doi{10.18653/v1/2026.acl-long.147}.
\newblock URL \url{https://doi.org/10.18653/v1/2026.acl-long.147}.

\bibitem[Kang and McAuley(2018)]{kang2018sasrec}
Wang-Cheng Kang and Julian McAuley.
\newblock Self-attentive sequential recommendation.
\newblock In \emph{2018 IEEE International Conference on Data Mining}, pages
  197--206. IEEE, 2018.
\newblock \doi{10.1109/ICDM.2018.00035}.
\newblock URL \url{https://doi.org/10.1109/ICDM.2018.00035}.

\bibitem[Kong et~al.(2025)Kong, Sheng, Tan, Chen, Wu, Zhang, Wang, and
  He]{MiniOneRec}
Xiaoyu Kong, Leheng Sheng, Junfei Tan, Yuxin Chen, Jiancan Wu, An~Zhang, Xiang
  Wang, and Xiangnan He.
\newblock {MiniOneRec}: An open-source framework for scaling generative
  recommendation, 2025.
\newblock URL \url{https://arxiv.org/abs/2510.24431}.

\bibitem[Lee et~al.(2024)Lee, Kang, Kweon, and Yu]{lee2024continual}
Gyuseok Lee, SeongKu Kang, Wonbin Kweon, and Hwanjo Yu.
\newblock Continual collaborative distillation for recommender system.
\newblock In \emph{Proceedings of the 30th ACM SIGKDD Conference on Knowledge
  Discovery and Data Mining}, pages 1495--1505. Association for Computing
  Machinery, 2024.
\newblock \doi{10.1145/3637528.3671924}.
\newblock URL \url{https://doi.org/10.1145/3637528.3671924}.

\bibitem[Lee et~al.(2019)Lee, Choi, Lee, and Shim]{lee2019collaborative}
Jae-Woong Lee, Minjin Choi, Jongwuk Lee, and Hyunjung Shim.
\newblock Collaborative distillation for {Top-N} recommendation.
\newblock In \emph{2019 IEEE International Conference on Data Mining}, pages
  369--378. IEEE, 2019.
\newblock \doi{10.1109/ICDM.2019.00047}.
\newblock URL \url{https://doi.org/10.1109/ICDM.2019.00047}.

\bibitem[Lee and Kim(2021)]{lee2021dual}
Youngjune Lee and Kee-Eung Kim.
\newblock Dual correction strategy for ranking distillation in {Top-N}
  recommender system.
\newblock In \emph{Proceedings of the 30th ACM International Conference on
  Information and Knowledge Management}, pages 3186--3190. Association for
  Computing Machinery, 2021.
\newblock \doi{10.1145/3459637.3482093}.
\newblock URL \url{https://doi.org/10.1145/3459637.3482093}.

\bibitem[Liao et~al.(2024)Liao, Li, Yang, Wu, Yuan, Wang, and
  He]{liao2024llara}
Jiayi Liao, Sihang Li, Zhengyi Yang, Jiancan Wu, Yancheng Yuan, Xiang Wang, and
  Xiangnan He.
\newblock {LLaRA}: Large language-recommendation assistant.
\newblock In \emph{Proceedings of the 47th International ACM SIGIR Conference
  on Research and Development in Information Retrieval}, pages 1785--1795.
  Association for Computing Machinery, 2024.
\newblock \doi{10.1145/3626772.3657690}.
\newblock URL \url{https://doi.org/10.1145/3626772.3657690}.

\bibitem[Lin et~al.(2025{\natexlab{a}})Lin, Gao, Chen, Zhou, Hu, Feng, Chen,
  and Wang]{lin2025spectral}
Siyi Lin, Chongming Gao, Jiawei Chen, Sheng Zhou, Binbin Hu, Yan Feng, Chun
  Chen, and Can Wang.
\newblock How do recommendation models amplify popularity bias? an analysis
  from the spectral perspective.
\newblock In \emph{Proceedings of the Eighteenth ACM International Conference
  on Web Search and Data Mining}, pages 659--668. Association for Computing
  Machinery, 2025{\natexlab{a}}.
\newblock \doi{10.1145/3701551.3703579}.
\newblock URL \url{https://doi.org/10.1145/3701551.3703579}.

\bibitem[Lin et~al.(2024)Lin, Wang, Li, Feng, Ng, and Chua]{lin2024transrec}
Xinyu Lin, Wenjie Wang, Yongqi Li, Fuli Feng, See-Kiong Ng, and Tat-Seng Chua.
\newblock Bridging items and language: A transition paradigm for large language
  model-based recommendation.
\newblock In \emph{Proceedings of the 30th ACM SIGKDD Conference on Knowledge
  Discovery and Data Mining}, pages 1816--1826. Association for Computing
  Machinery, 2024.
\newblock \doi{10.1145/3637528.3671884}.
\newblock URL \url{https://doi.org/10.1145/3637528.3671884}.

\bibitem[Lin et~al.(2025{\natexlab{b}})Lin, Shi, Wang, Feng, Wang, Ng, and
  Chua]{lin2025setrec}
Xinyu Lin, Haihan Shi, Wenjie Wang, Fuli Feng, Qifan Wang, See-Kiong Ng, and
  Tat-Seng Chua.
\newblock Order-agnostic identifier for large language model-based generative
  recommendation.
\newblock In \emph{Proceedings of the 48th International ACM SIGIR Conference
  on Research and Development in Information Retrieval}, pages 1923--1933.
  Association for Computing Machinery, 2025{\natexlab{b}}.
\newblock \doi{10.1145/3726302.3730053}.
\newblock URL \url{https://doi.org/10.1145/3726302.3730053}.

\bibitem[Liu et~al.(2020)Liu, Cheng, Dong, He, Pan, and
  Ming]{liu2020counterfactual}
Dugang Liu, Pengxiang Cheng, Zhenhua Dong, Xiuqiang He, Weike Pan, and Zhong
  Ming.
\newblock A general knowledge distillation framework for counterfactual
  recommendation via uniform data.
\newblock In \emph{Proceedings of the 43rd International ACM SIGIR Conference
  on Research and Development in Information Retrieval}, pages 831--840.
  Association for Computing Machinery, 2020.
\newblock \doi{10.1145/3397271.3401083}.
\newblock URL \url{https://doi.org/10.1145/3397271.3401083}.

\bibitem[Ma et~al.(2026)Ma, Wei, Zhao, Zhang, Xiao, Li, Yang, Gao, Wang, Li,
  et~al.]{ma2026mopd}
Wenhan Ma, Jianyu Wei, Liang Zhao, Hailin Zhang, Bangjun Xiao, Lei Li, Qibin
  Yang, Bofei Gao, Yudong Wang, Rang Li, et~al.
\newblock {MOPD}: Multi-teacher on-policy distillation for capability
  integration in {LLM} post-training, 2026.
\newblock URL \url{https://arxiv.org/abs/2606.30406}.

\bibitem[{OneRec Team} et~al.(2026){OneRec Team}, Yang, Ding, Chu, Zang, Pan,
  Li, Jiang, Bao, Wang, et~al.]{onerec2026onereason}
{OneRec Team}, Biao Yang, Boyang Ding, Chenglong Chu, Dunju Zang, Fei Pan, Han
  Li, Hao Jiang, Honghui Bao, Huanjie Wang, et~al.
\newblock {OneReason} technical report, 2026.
\newblock URL \url{https://arxiv.org/abs/2606.06260}.

\bibitem[Qi et~al.(2026)Qi, Zou, Yang, Xu, Sun, and Li]{qi2026ampo}
Junbo Qi, Yanyan Zou, Xuanhua Yang, Sulong Xu, Ying Sun, and Shengjie Li.
\newblock Adaptive mix preference optimization for generative recommendation.
\newblock In \emph{Proceedings of the 49th International ACM SIGIR Conference
  on Research and Development in Information Retrieval}, pages 1485--1495.
  Association for Computing Machinery, 2026.
\newblock \doi{10.1145/3805712.3809727}.
\newblock URL \url{https://doi.org/10.1145/3805712.3809727}.

\bibitem[Rajput et~al.(2023)Rajput, Mehta, Singh, Keshavan, Vu, Heldt, Hong,
  Tay, Tran, Samost, et~al.]{rajput2023recommender}
Shashank Rajput, Nikhil Mehta, Anima Singh, Raghunandan~H. Keshavan, Trung Vu,
  Lukasz Heldt, Lichan Hong, Yi~Tay, Vinh~Q. Tran, Jonah Samost, et~al.
\newblock Recommender systems with generative retrieval.
\newblock In \emph{Advances in Neural Information Processing Systems},
  volume~36, pages 10299--10315, 2023.
\newblock \doi{10.52202/075280-0452}.
\newblock URL \url{https://doi.org/10.52202/075280-0452}.

\bibitem[Saito et~al.(2020)Saito, Yaginuma, Nishino, Sakata, and
  Nakata]{saito2020unbiased}
Yuta Saito, Suguru Yaginuma, Yuta Nishino, Hayato Sakata, and Kazuhide Nakata.
\newblock Unbiased recommender learning from missing-not-at-random implicit
  feedback.
\newblock In \emph{Proceedings of the 13th International Conference on Web
  Search and Data Mining}, pages 501--509. Association for Computing Machinery,
  2020.
\newblock \doi{10.1145/3336191.3371783}.
\newblock URL \url{https://doi.org/10.1145/3336191.3371783}.

\bibitem[Tan et~al.(2024)Tan, Xu, Hua, Ge, Li, and Zhang]{tan2024idgenrec}
Juntao Tan, Shuyuan Xu, Wenyue Hua, Yingqiang Ge, Zelong Li, and Yongfeng
  Zhang.
\newblock {IDGenRec}: {LLM}-recsys alignment with textual {ID} learning.
\newblock In \emph{Proceedings of the 47th International ACM SIGIR Conference
  on Research and Development in Information Retrieval}, pages 355--364.
  Association for Computing Machinery, 2024.
\newblock \doi{10.1145/3626772.3657821}.
\newblock URL \url{https://doi.org/10.1145/3626772.3657821}.

\bibitem[Tang and Wang(2018)]{tang2018ranking}
Jiaxi Tang and Ke~Wang.
\newblock Ranking distillation: Learning compact ranking models with high
  performance for recommender system.
\newblock In \emph{Proceedings of the 24th ACM SIGKDD International Conference
  on Knowledge Discovery and Data Mining}, pages 2289--2298. Association for
  Computing Machinery, 2018.
\newblock \doi{10.1145/3219819.3220021}.
\newblock URL \url{https://doi.org/10.1145/3219819.3220021}.

\bibitem[Wang et~al.(2021{\natexlab{a}})Wang, Feng, He, Nie, and
  Chua]{wang2021denoising}
Wenjie Wang, Fuli Feng, Xiangnan He, Liqiang Nie, and Tat-Seng Chua.
\newblock Denoising implicit feedback for recommendation.
\newblock In \emph{Proceedings of the 14th ACM International Conference on Web
  Search and Data Mining}, pages 373--381. Association for Computing Machinery,
  2021{\natexlab{a}}.
\newblock \doi{10.1145/3437963.3441800}.
\newblock URL \url{https://doi.org/10.1145/3437963.3441800}.

\bibitem[Wang et~al.(2021{\natexlab{b}})Wang, Feng, He, Wang, and
  Chua]{wang2021deconfounded}
Wenjie Wang, Fuli Feng, Xiangnan He, Xiang Wang, and Tat-Seng Chua.
\newblock Deconfounded recommendation for alleviating bias amplification.
\newblock In \emph{Proceedings of the 27th ACM SIGKDD Conference on Knowledge
  Discovery and Data Mining}, pages 1717--1725. Association for Computing
  Machinery, 2021{\natexlab{b}}.
\newblock \doi{10.1145/3447548.3467249}.
\newblock URL \url{https://doi.org/10.1145/3447548.3467249}.

\bibitem[Wang et~al.(2024{\natexlab{a}})Wang, Bao, Lin, Zhang, Li, Feng, Ng,
  and Chua]{wang2024letter}
Wenjie Wang, Honghui Bao, Xinyu Lin, Jizhi Zhang, Yongqi Li, Fuli Feng,
  See-Kiong Ng, and Tat-Seng Chua.
\newblock Learnable item tokenization for generative recommendation.
\newblock In \emph{Proceedings of the 33rd ACM International Conference on
  Information and Knowledge Management}, pages 2400--2409. Association for
  Computing Machinery, 2024{\natexlab{a}}.
\newblock \doi{10.1145/3627673.3679569}.
\newblock URL \url{https://doi.org/10.1145/3627673.3679569}.

\bibitem[Wang et~al.(2024{\natexlab{b}})Wang, Xun, Hong, Zhu, Jin, Lin, Li, Li,
  Xia, Zhao, et~al.]{wang2024eager}
Ye~Wang, Jiahao Xun, Minjie Hong, Jieming Zhu, Tao Jin, Wang Lin, Haoyuan Li,
  Linjun Li, Yan Xia, Zhou Zhao, et~al.
\newblock {EAGER}: Two-stream generative recommender with behavior-semantic
  collaboration.
\newblock In \emph{Proceedings of the 30th ACM SIGKDD Conference on Knowledge
  Discovery and Data Mining}, pages 3245--3254. Association for Computing
  Machinery, 2024{\natexlab{b}}.
\newblock \doi{10.1145/3637528.3671775}.
\newblock URL \url{https://doi.org/10.1145/3637528.3671775}.

\bibitem[Wei et~al.(2021)Wei, Feng, Chen, Wu, Yi, and He]{wei2021macr}
Tianxin Wei, Fuli Feng, Jiawei Chen, Ziwei Wu, Jinfeng Yi, and Xiangnan He.
\newblock Model-agnostic counterfactual reasoning for eliminating popularity
  bias in recommender system.
\newblock In \emph{Proceedings of the 27th ACM SIGKDD Conference on Knowledge
  Discovery and Data Mining}, pages 1791--1800. Association for Computing
  Machinery, 2021.
\newblock \doi{10.1145/3447548.3467289}.
\newblock URL \url{https://doi.org/10.1145/3447548.3467289}.

\bibitem[Xv et~al.(2022)Xv, Lin, Li, Su, Ye, and Chen]{xv2022neutralizing}
Guipeng Xv, Chen Lin, Hui Li, Jinsong Su, Weiyao Ye, and Yewang Chen.
\newblock Neutralizing popularity bias in recommendation models.
\newblock In \emph{Proceedings of the 45th International ACM SIGIR Conference
  on Research and Development in Information Retrieval}, pages 2623--2628.
  Association for Computing Machinery, 2022.
\newblock \doi{10.1145/3477495.3531907}.
\newblock URL \url{https://doi.org/10.1145/3477495.3531907}.

\bibitem[Yang et~al.(2026)Yang, Wang, Xu, Chen, Zhang, Chen, Jin, and
  Wang]{yang2026bear}
Weiqin Yang, Bohao Wang, Zhenxiang Xu, Jiawei Chen, Shengjia Zhang, Jingbang
  Chen, Canghong Jin, and Can Wang.
\newblock {BEAR}: Towards beam-search-aware optimization for recommendation
  with large language models.
\newblock In \emph{Proceedings of the 49th International ACM SIGIR Conference
  on Research and Development in Information Retrieval}, pages 2196--2207.
  Association for Computing Machinery, 2026.
\newblock \doi{10.1145/3805712.3809533}.
\newblock URL \url{https://doi.org/10.1145/3805712.3809533}.

\bibitem[Yin et~al.(2026)Yin, Zhu, Huo, Liu, Chen, Wang, Pan, and
  Zhang]{yin2026ghost}
Jun Yin, Bangguo Zhu, Peng Huo, Ruochen Liu, Hao Chen, Senzhang Wang, Shirui
  Pan, and Chengqi Zhang.
\newblock Echoes in filter bubble: Diagnosing and curing popularity bias in
  generative recommender systems.
\newblock In \emph{NeurIPS}, 2026.
\newblock URL \url{https://neurips.cc/virtual/2026/poster/148459}.
\newblock Accepted; proceedings forthcoming.

\bibitem[Zhang and Shen(2023)]{zhang2023popularity}
Fan Zhang and Qijie Shen.
\newblock A model-agnostic popularity debias training framework for
  click-through rate prediction in recommender system.
\newblock In \emph{Proceedings of the 46th International ACM SIGIR Conference
  on Research and Development in Information Retrieval}, pages 1760--1764.
  Association for Computing Machinery, 2023.
\newblock \doi{10.1145/3539618.3591939}.
\newblock URL \url{https://doi.org/10.1145/3539618.3591939}.

\bibitem[Zhang et~al.(2021)Zhang, Feng, He, Wei, Song, Ling, and
  Zhang]{zhang2021pda}
Yang Zhang, Fuli Feng, Xiangnan He, Tianxin Wei, Chonggang Song, Guohui Ling,
  and Yongdong Zhang.
\newblock Causal intervention for leveraging popularity bias in recommendation.
\newblock In \emph{Proceedings of the 44th International ACM SIGIR Conference
  on Research and Development in Information Retrieval}, pages 11--20.
  Association for Computing Machinery, 2021.
\newblock \doi{10.1145/3404835.3462875}.
\newblock URL \url{https://doi.org/10.1145/3404835.3462875}.

\bibitem[Zheng et~al.(2021)Zheng, Gao, Li, He, Li, and Jin]{zheng2021dice}
Yu~Zheng, Chen Gao, Xiang Li, Xiangnan He, Yong Li, and Depeng Jin.
\newblock Disentangling user interest and conformity for recommendation with
  causal embedding.
\newblock In \emph{Proceedings of the Web Conference 2021}, pages 2980--2991.
  Association for Computing Machinery, 2021.
\newblock \doi{10.1145/3442381.3449788}.
\newblock URL \url{https://doi.org/10.1145/3442381.3449788}.

\bibitem[Zhu et~al.(2021)Zhu, He, Zhao, and Caverlee]{zhu2021dynamic}
Ziwei Zhu, Yun He, Xing Zhao, and James Caverlee.
\newblock Popularity bias in dynamic recommendation.
\newblock In \emph{Proceedings of the 27th ACM SIGKDD Conference on Knowledge
  Discovery and Data Mining}, pages 2439--2449. Association for Computing
  Machinery, 2021.
\newblock \doi{10.1145/3447548.3467376}.
\newblock URL \url{https://doi.org/10.1145/3447548.3467376}.

\bibitem[Zhuang et~al.(2025)Zhuang, Du, Han, Li, Fu, Jose, and
  Ni]{zhuang2025wasserstein}
Ziyi Zhuang, Hanwen Du, Hui Han, Youhua Li, Junchen Fu, Joemon~M. Jose, and
  Yongxin Ni.
\newblock Bridging the gap: Teacher-assisted wasserstein knowledge distillation
  for efficient multi-modal recommendation.
\newblock In \emph{Proceedings of the ACM Web Conference 2025}, pages
  2464--2475. Association for Computing Machinery, 2025.
\newblock \doi{10.1145/3696410.3714852}.
\newblock URL \url{https://doi.org/10.1145/3696410.3714852}.

\end{thebibliography}

\clearpage
\appendix
\section{Public-Dataset Ablation Results}
Table~\ref{tab:public-ablation} shows the results of ablations on public datasets. HERO achieves the best overall performance across the three datasets.
\label{app:public-ablation}

\begin{table}[H]
\centering
\caption{Public-dataset component ablation (Hit ratios).
Tail@20 denotes Tail Hit@20.}
\label{tab:public-ablation}
\scriptsize
\setlength{\tabcolsep}{1.5pt}
\renewcommand{\arraystretch}{1.08}
\begin{tabular*}{\columnwidth}{@{\extracolsep{\fill}}lrrrrrr@{}}
\toprule
& \multicolumn{2}{c}{Beauty}
& \multicolumn{2}{c}{Pet}
& \multicolumn{2}{c}{MovieLens-1M} \\
\cmidrule(lr){2-3}
\cmidrule(lr){4-5}
\cmidrule(lr){6-7}
Variant
& Hit@20 & Tail@20
& Hit@20 & Tail@20
& Hit@20 & Tail@20 \\
\midrule
\rowcolor{oursgray}
\textbf{Ours}
& 0.0391 & 0.0127 & 0.0384 & 0.0144 & 0.1936 & 0.0995 \\
w/o rollout KL
& 0.0378 & 0.0107 & 0.0381 & 0.0141 & 0.1721 & 0.0873 \\
Head rollout teacher
& 0.0383 & 0.0021 & 0.0369 & 0.0037 & 0.1709 & 0.0538 \\
Tail rollout teacher
& 0.0375 & 0.0105 & 0.0347 & 0.0146 & 0.1741 & 0.0889 \\
50/50 rollout mixture
& 0.0366 & 0.0065 & 0.0376 & 0.0082 & 0.1756 & 0.0705 \\
w/o GT-CE
& 0.0362 & 0.0088 & 0.0352 & 0.0121 & 0.1751 & 0.0859 \\
w/o tail-path GT-KL
& 0.0384 & 0.0095 & 0.0382 & 0.0109 & 0.1768 & 0.0776 \\
\bottomrule
\end{tabular*}
\end{table}

\section{Accuracy Gains across User Activity Strata}
\label{app:user-activity}

We group evaluation users on each public dataset by their training interaction counts. Users without training history form the New group, while the remaining users are divided into activity quartiles. Adjacent low-support strata are merged to ensure at least 100 SFT hits per retained stratum.

\begin{figure}[!htbp]
    \centering
    \includegraphics[width=\columnwidth]{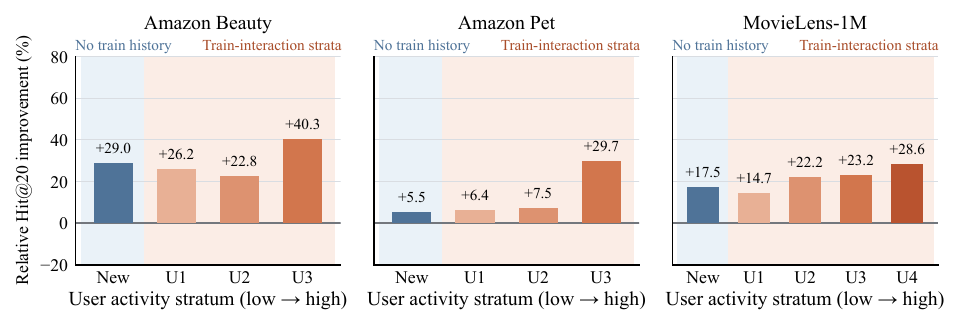}
    \caption{Relative Hit@20 improvement over SFT by user activity stratum. New denotes users without training history; U1 onward order users with training history from lower to higher activity.}
    \label{fig:user-activity-gain}
\end{figure}

As shown in Figure~\ref{fig:user-activity-gain}, \methodname consistently improves Hit@20 across all retained activity strata on all three datasets. These gains extend from users without training history to highly active users, demonstrating its effectiveness across different levels of user activity.

\section{Transfer of Teacher Specialization}
\label{app:teacher-transfer}

\begin{figure}[!htbp]
    \centering
    \includegraphics[width=\columnwidth]{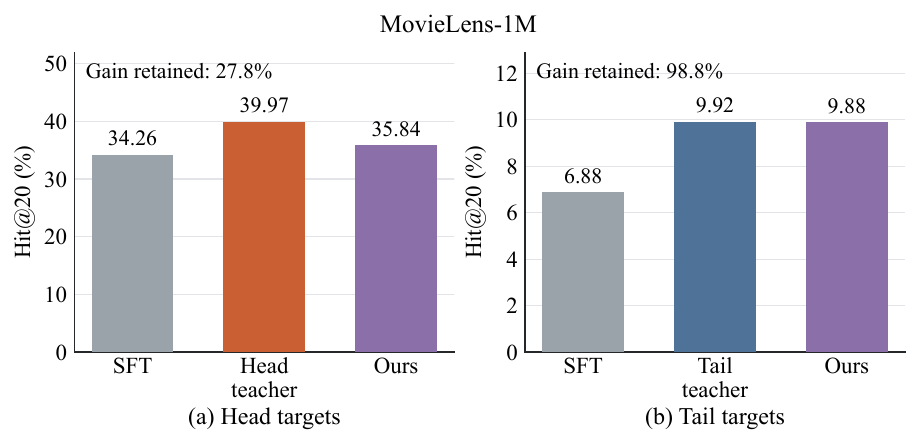}
    \caption{Hit@20 on head and tail targets in the MovieLens-1M diagnostic cohort. Our method improves over SFT in both groups and nearly matches the tail teacher.}
    \label{fig:teacher-transfer}
\end{figure}

We examine teacher specialization transfer on the first 4,096 chronologically ordered MovieLens-1M validation contexts using the same trie-constrained Beam-20 decoding. Gain retained measures the student's improvement over SFT relative to the corresponding teacher's improvement. As shown in Figure~\ref{fig:teacher-transfer}, \methodname improves both head and tail accuracy within a single student, nearly matching the tail teacher and retaining almost all of its gain on tail targets.

\section{Baseline Details}
\label{app:baseline}

All generative baselines use the same fixed SIDs, candidate trie, evaluation
split, and decoding as HERO, and all distillation baselines start from the
same SFT checkpoint $p_{\theta_0}$, use the same post-training data, are
trained for the same number of steps as HERO, and use forward KL. Learning
rates are searched over $\{10^{-5}, 5\times10^{-5}, 10^{-4}\}$, and the
remaining hyperparameters follow the original papers.
\emph{SASRec}~\cite{kang2018sasrec} and
\emph{TIGER}~\cite{rajput2023recommender} follow the model structures in
MiniOneRec~\cite{MiniOneRec}, with TIGER trained on our fixed SIDs.
\emph{SFT++} continues from $p_{\theta_0}$ with the CE loss on the
post-training data. \emph{GT-KD} uses HERO's routed teacher $q_g$ to
supervise the ground-truth prefixes of all examples. \emph{Off-policy KD}
and \emph{GKD}~\cite{agarwal2024onpolicy} use a single teacher
$q_{\mathrm{all}}$ trained on the full $\mathcal{D}_{\mathrm{SFT}}$ with the
same architecture and recipe as $q_h$ and $q_t$, and distill on the top-$J$
beams ($J = 32$, as in HERO) of the frozen $p_{\theta_0}$ and of the current
student, respectively, together with the GT-CE loss; GKD is thus the
standard OPD objective in Eq.~\eqref{eq:opd} plus the GT-CE loss.
\emph{MOPD}~\cite{ma2026mopd} routes each example to $q_h$ or $q_t$ and
distills on the student's own rollouts. \emph{SID-GRPO} follows MiniOneRec
with 32 samples per context, and \emph{Ghost-AUO}~\cite{yin2026ghost} and
\emph{BEAR}~\cite{yang2026bear} follow their original implementations.

\end{document}